\documentclass[a4paper,11pt]{article}
\usepackage{pos}

\title{Quark mass dependence of doubly heavy tetraquark binding}

\author*[a]{W. G. Parrott}
\author[b]{B. Colquhoun}
\author[c]{A. Francis}
\author[d]{R. J. Hudspith}
\author[a]{R. Lewis}
\author[e,f]{K. Maltman}

\affiliation[a]{Department of Physics and Astronomy,\\
  York University, Toronto, Ontario, M3J 1P3, Canada}

\affiliation[b]{SUPA, School of Physics and Astronomy, University of Glasgow,\\
  Glasgow, G12 8QQ, UK}

\affiliation[c]{Institute of Physics, National Yang Ming Chiao Tung University,\\
  30010 Hsinchu, Taiwan}
\affiliation[d]{GSI Helmholtzzentrum für Schwerionenforschung,\\
  64291 Darmstadt, Germany}
\affiliation[e]{Department of Mathematics and Statistics, York University,\\
  Toronto, Ontario M3J 1P3, Canada}
\affiliation[f]{CSSM, University of Adelaide,\\
  Adelaide, SA, 5005, Australia}

\emailAdd{parrott@yorku.ca}
\emailAdd{brian.colquhoun@glasgow.ac.uk}
\emailAdd{afrancis.heplat@googlemail.com}
\emailAdd{renwick.james.hudspith@googlemail.com}
\emailAdd{randy.lewis@yorku.ca}
\emailAdd{kmaltman@yorku.ca}
\abstract{The existence of bound doubly heavy tetraquark states was confirmed by the recent LHCb discovery of the doubly charmed $T_{cc}$, less than 1 MeV below the meson pair threshold. Others states with two heavy (bottom or charm) quarks could also be bound, perhaps more deeply. Here we discuss our previous work, and the improvements in our current, updated analysis of various heavy-heavy-light-light tetraquark candidates, including the light and heavy quark mass dependence of the binding.}

\FullConference{The 41st International Symposium on Lattice Field Theory (LATTICE2024)\\
 28 July - 3 August 2024\\
Liverpool, UK\\}

\begin{document}
\maketitle

\section{Introduction}
The existence of strong-interaction-stable doubly heavy tetraquarks, $q_1q_2\bar{h}\bar{h}$ (where $h$ are heavy quarks with $m_h\geq m_c$  and $q_1$, $q_2$ are the light quark flavours respectively) in the heavy mass limit $m_h\to\infty$ has long been suspected~\cite{Ader:1981db,Heller:1986bt,Manohar:1992nd}. Phenomenologically, such states have access to a binding contribution from the colour Coulomb interaction between heavy antiquarks in a colour $3_c$ configuration, which is proportional to the reduced mass of the heavy antiquark pair. Additional binding contributions are present for $J^P=1^+$ or $J^P=0^+$ heavy-heavy-light-light tetraquarks, where the light quarks are either $I=0$ ($ud$) or $I=1/2$ ($ls$, with $l\in\{u, d\}$). In this case, the light degrees of freedom are in the flavor anti-symmetric, colour $\bar{3}_c$, light-quark spin $J_l =0$ ``good light diquark'' configuration, known to be attractive from the observed splittings in the heavy baryon spectrum, and lattice calculations~\cite{Francis:2021vrr}.

To what extent this binding survives to physical heavy masses $m_b\geq m_h\geq m_c$ remains an active field of study on the lattice~\cite{Bicudo:2015kna,Bicudo:2016ooe,Bicudo:2017szl,Francis:2016hui,Junnarkar:2018twb,Leskovec:2019ioa,Mohanta:2020eed,Hudspith:2020tdf,Pflaumer:2020ogv,Pflaumer:2021ong,Wagner:2022bff,Pflaumer:2022lgp,Colquhoun:2022dte,Colquhoun:2022sip,Aoki:2023nzp,Hudspith:2023loy,Mueller:2023wzd,Alexandrou:2024iwi,Meinel:2022lzo}. In the case of $ud\bar{c}\bar{c}$, a 2021 LHCb measurement~\cite{LHCb:2021vvq,LHCb:2021auc} found a bound $J^P=1^+$ state lying less that 1 MeV below the $DD^*$ threshold. This binding is well below the precision of current lattice calculations, but the finding strongly implies binding in similar channels with heavier heavy quarks, such as $ud\bar{b}\bar{b}$. 

These proceedings concern our recent work~\cite{Colquhoun:2024jzh}, containing an updated analysis of the $ud\bar{h}\bar{h}$, $ls\bar{h}\bar{h}$, $ud\bar{b}\bar{h}$, and $ls\bar{b}\bar{h}$ $J^P=1^+$ and $J^P=0^+$ tetraquark states, in the context of the previous work by the collaboration~\cite{Francis:2016hui,Francis:2018jyb,Hudspith:2020tdf}. We shall begin by briefly summarising the previous work, before moving onto the changes and updates made in the present analysis, and the results.
\section{Previous work}
The work in~\cite{Colquhoun:2024jzh} is the culmination of previous studies~\cite{Francis:2016hui,Francis:2018jyb,Hudspith:2020tdf}, which investigated both the light and heavy mass dependence of the tetraquark binding, as well as a number of different tetraquark channels.
Both the previous work and the current update employ PACS-CS $N_f=2+1$ Wilson-Clover ensembles~\cite{Aoki:2009ix}, with $a^{-1}=2.194(10)$GeV~\cite{PACS-CS:2011ngu}, with Ref.~\cite{Hudspith:2020tdf} and the current update adding new ensembles supplementing the PACS-CS set. Non-Relativistic QCD (NRQCD) is used to reach heavy masses above $m_c$.
\subsection{Light quark mass dependence}
The collaboration's first steps towards our present calculation were taken in~\cite{Francis:2016hui}. This focused on the $J^P=1^+$ $ud\bar{b}\bar{b}$ and $ls\bar{b}\bar{b}$ tetraquarks, with three values for the pion mass in the calculation ($M_{\pi}=\in\{165,299,415\}$ MeV). Two different operators which should overlap with the tetraquark state were used; a meson-meson like operator,
\begin{equation}
  M(x)=\bar{b}_a^{\alpha}(x)\gamma_5^{\alpha\beta}u^{\beta}_{a}(x)\bar{b}_b^{\kappa}(x)\gamma_i^{\kappa\rho}u^{\rho}_b(x) - u\leftrightarrow d,
\end{equation}
and a diquark-antidiquark like operator,
\begin{equation}
  D(x)=\big(u^{\alpha}_a(x)\big)^T\big(C\gamma_5\big)^{\alpha\beta}d^{\beta}_b(x)\bar{b}^{\kappa}_{a}(x)\big(C\gamma_i\big)^{\kappa\rho}\big(\bar{b}^{\rho}_b(x)\big)^T,
\end{equation}
where $C$ is the charge conjugation operator.

The two-point correlation function corresponding to each combination of $M$ and $D$ at source and sink, $C_{\mathcal{O}_1\mathcal{O}_2}(t)$, is combined with the two-point correlation functions for the threshold pseudoscalar ($C_{PP}(t)$) and vector ($C_{VV}(t)$) meson states to create a matrix for use in a Generalised Eigenvalue Problem (GEVP),
\begin{equation}
  F(t)=\frac{1}{C_{PP}(t)C_{VV}(t)}
  \begin{pmatrix}
    C_{DD}(t)&C_{DM}(t)\\
    C_{MD}(t)&C_{MM}(t)\\
  \end{pmatrix},
\end{equation}
which has a lowest Eigenvalue given by $\lambda_1$:
\begin{equation}\label{eq:GEVP}
  \begin{split}
    F(t)\nu&=\lambda(t)F(t_0)\nu\\
    \lambda_1(t) &= Ae^{-\Delta E(t-t_0)},
  \end{split}
\end{equation}
which can be fitted to extract the splitting between the tetraquark and threshold state $\Delta E$.

\begin{figure}
  \begin{center}
    \includegraphics[width=0.48\textwidth]{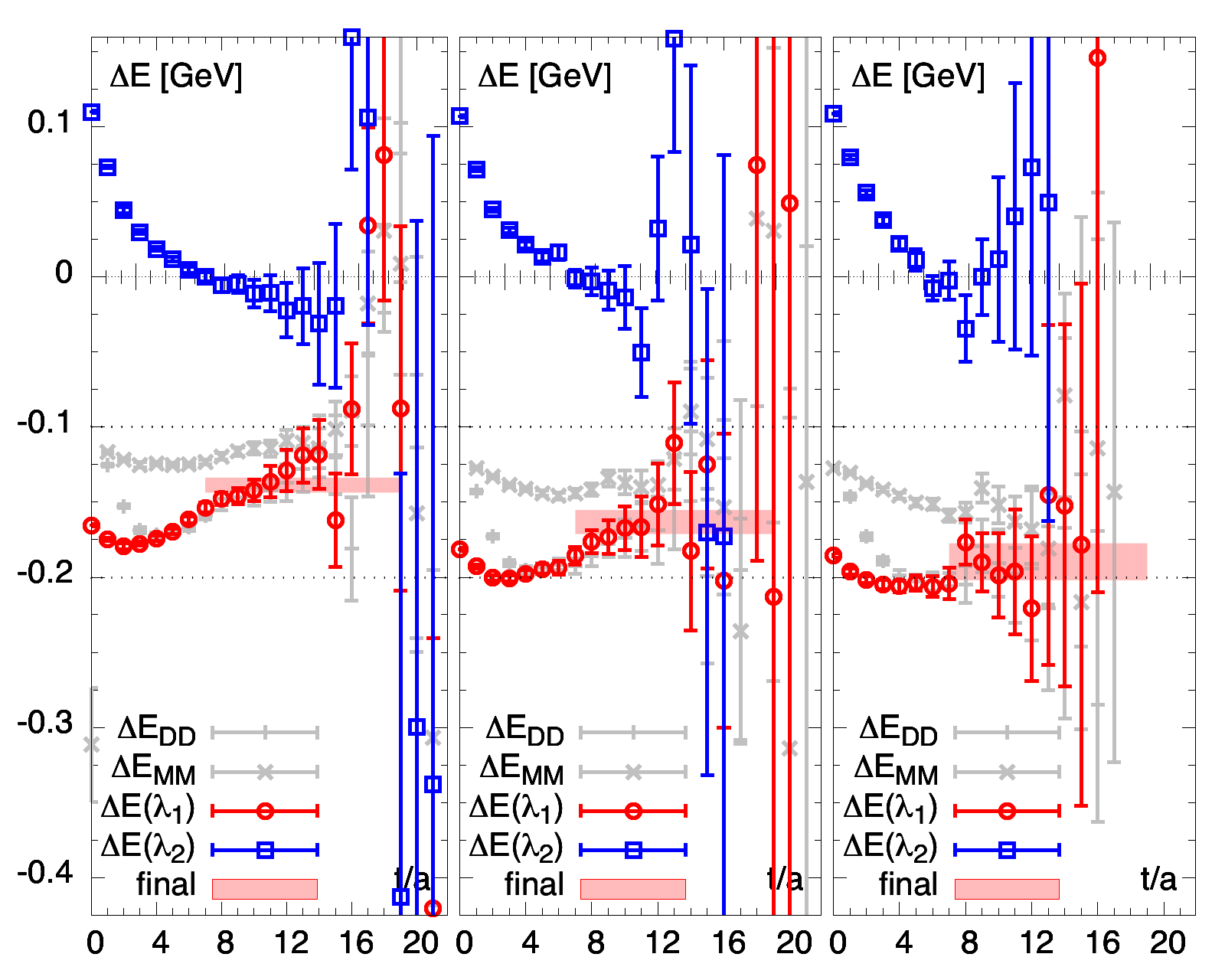}
    \caption{Taken from~\cite{Francis:2016hui}, $\Delta E$ (for $ud\bar{b}\bar{b}$) extracted from Eq.~\eqref{eq:GEVP}, from different values of $t$, for the two Eigenvalues $\lambda_1$ (red) and $\lambda_2$ (blue). The final value taken for $\Delta E(\lambda_1)$ is shown as a red band.}
    \label{fig:2016_plat}
  \end{center}
\end{figure}

Fig.~\ref{fig:2016_plat} is taken from~\cite{Francis:2016hui}, and shows the resulting $\Delta E$ values for different $t$ choices in red, as well as the final value (red band). These results show plateaus which appear to rise at large $t$ values.
\begin{figure}
  \begin{center}
    \includegraphics[width=0.48\textwidth]{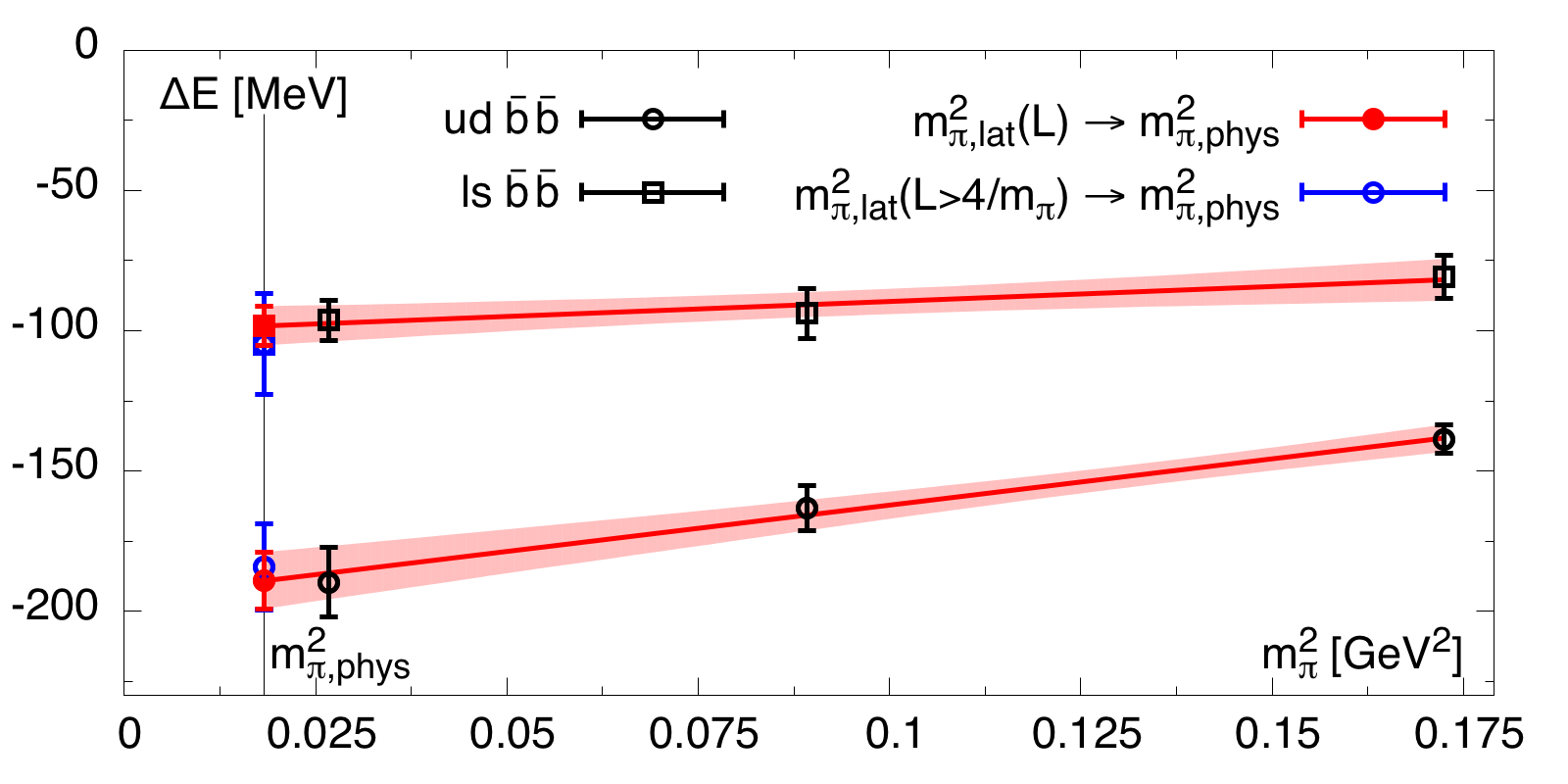}
    \caption{Taken from~\cite{Francis:2016hui}, $ud\bar{b}\bar{b}$ and $ls\bar{b}\bar{b}$ for the three different $M_{\pi}$ values, with a linear extrapolation in $M_{\pi}^2$ shown in red. The red and blue points give the extrapolated value at physical pion mass, the blue one excluding the leftmost data point, for which $M_{\pi}L<4$.}
    \label{fig:2016_chiral}
  \end{center}
\end{figure}    
Taking the binding results for the three different ensembles used, a linear fit in $M_{\pi}^2$ is extrapolated to the physical pion mass. This fit is shown in Fig.~\ref{fig:2016_chiral}, taken from~\cite{Francis:2016hui}. The fits both appear linear, with stronger pion mass dependence visible in the $ud\bar{b}\bar{b}$ case. Physical pion mass binding energies of $189(10)$ MeV and $98(7)$ MeV are quoted for the $ud\bar{b}\bar{b}$ and $ls\bar{b}\bar{b}$ tetraquarks respectively. These results suggest that the tetraquarks are deeply bound relative to the thresholds, and whilst no finite volume (FV) analysis is carried out in this paper, the depth of this binding is greater than any expected shift from FV effects. This is further supported by the presence of an excited state close to threshold.     
\subsection{Heavy quark mass dependence}
The next steps were taken in~\cite{Francis:2018jyb}. This work used one ensemble from the previous work with $M_{\pi}=299$ MeV, and again focused on the $J^p=1^+$ tetraquarks. This time, a variable heavy mass was employed, with $am_h\in\{0.9,1.0,1.2,1.6,3.0,4.0,8.0,10.0\}$, giving $0.1\leq\frac{m_b}{m_h}\leq1.8$, and four tetraquark channels were studied: $ud\bar{b}\bar{h}$, $ud\bar{h}\bar{h}$, $ls\bar{b}\bar{h}$ and $ls\bar{h}\bar{h}$. The analysis proceeds as in~\cite{Francis:2016hui} above, but in the $q_1q_2\bar{b}\bar{h}$ case, a $3\times3$ matrix is permitted for the GEVP by the exchange $b\leftrightarrow h$ in $D(x)$. Higher order Eigenvalues give access to excited states.
\begin{figure}
  \begin{center}
    \includegraphics[width=0.48\textwidth]{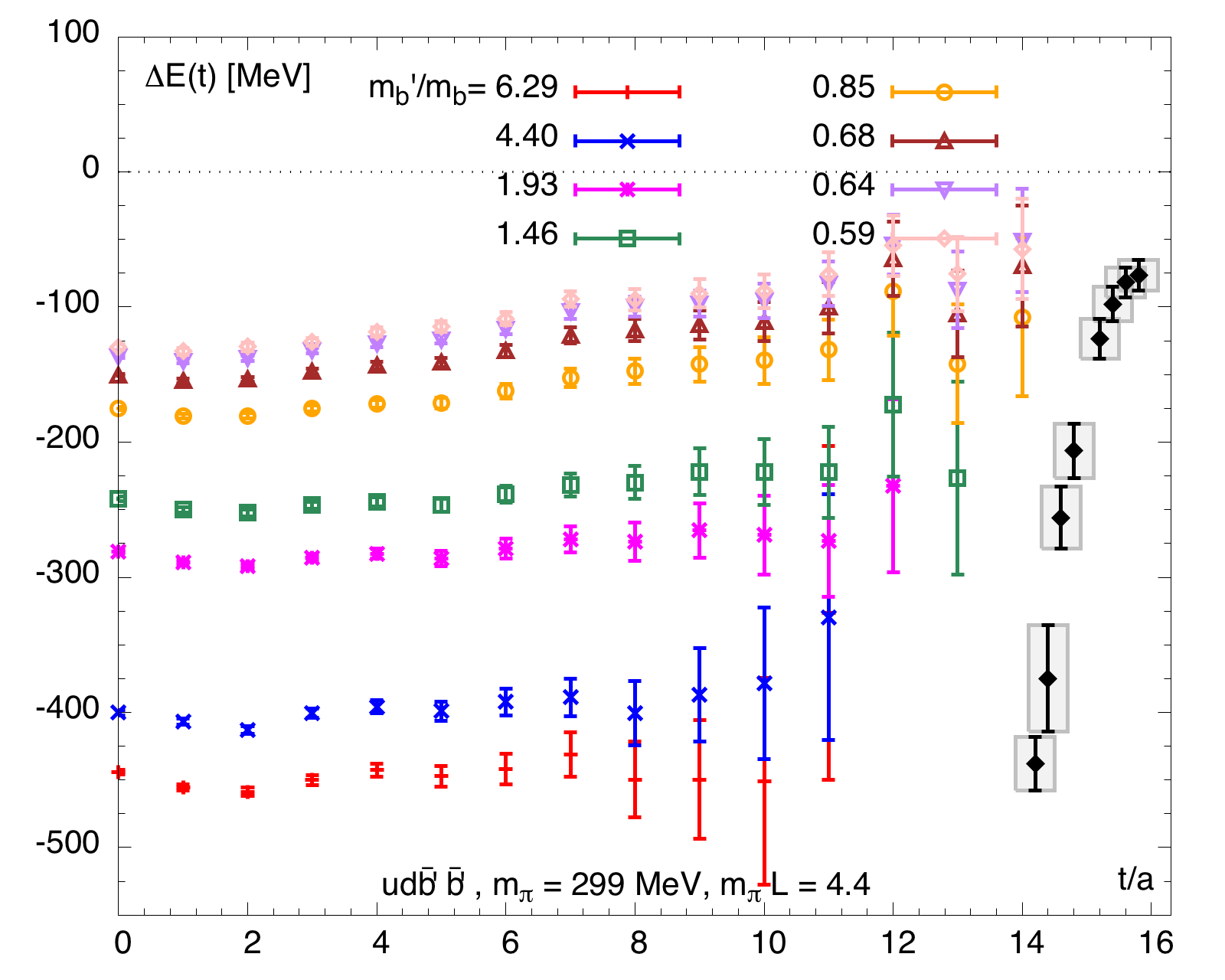}
    \includegraphics[width=0.48\textwidth]{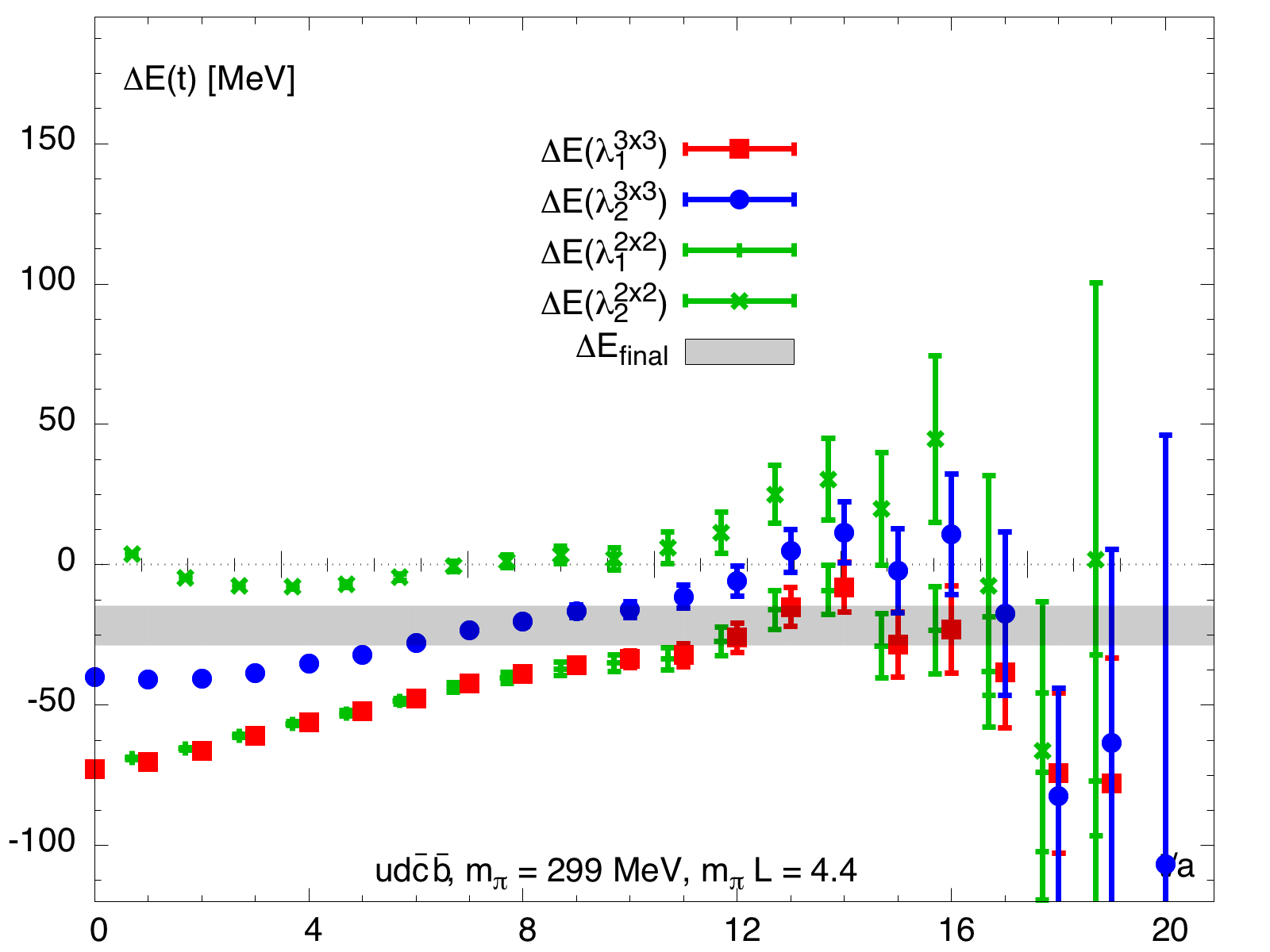}
    \caption{Taken from~\cite{Francis:2018jyb}. Left: $\Delta E$ (for $ud\bar{h}\bar{h}$, where the notation $b'\equiv h$ is used) extracted from Eq.~\eqref{eq:GEVP}, from different values of $t$ and choices of $m_h$, for the Eigenvalue $\lambda_1$. The final value taken for $\Delta E(\lambda_1)$ is shown as a black point. Right: $\Delta E$ plateaus for $ud\bar{b}\bar{c}$, showing $\lambda_1$ and $\lambda_2$ for the $2\times2$ and $3\times3$ GEVP cases.}
    \label{fig:2018_plat}
  \end{center}
\end{figure}
The $\Delta E$ plateaus in~\cite{Francis:2018jyb} shown in Fig.~\ref{fig:2018_plat}, particularly for $ud\bar{b}\bar{c}$, exhibit similar behaviour to those in~\cite{Francis:2016hui}, with it hard to determine whether or not they are still rising when the noise begins to dominate at large $t$ values.
\begin{figure}
  \begin{center}
    \includegraphics[width=0.48\textwidth]{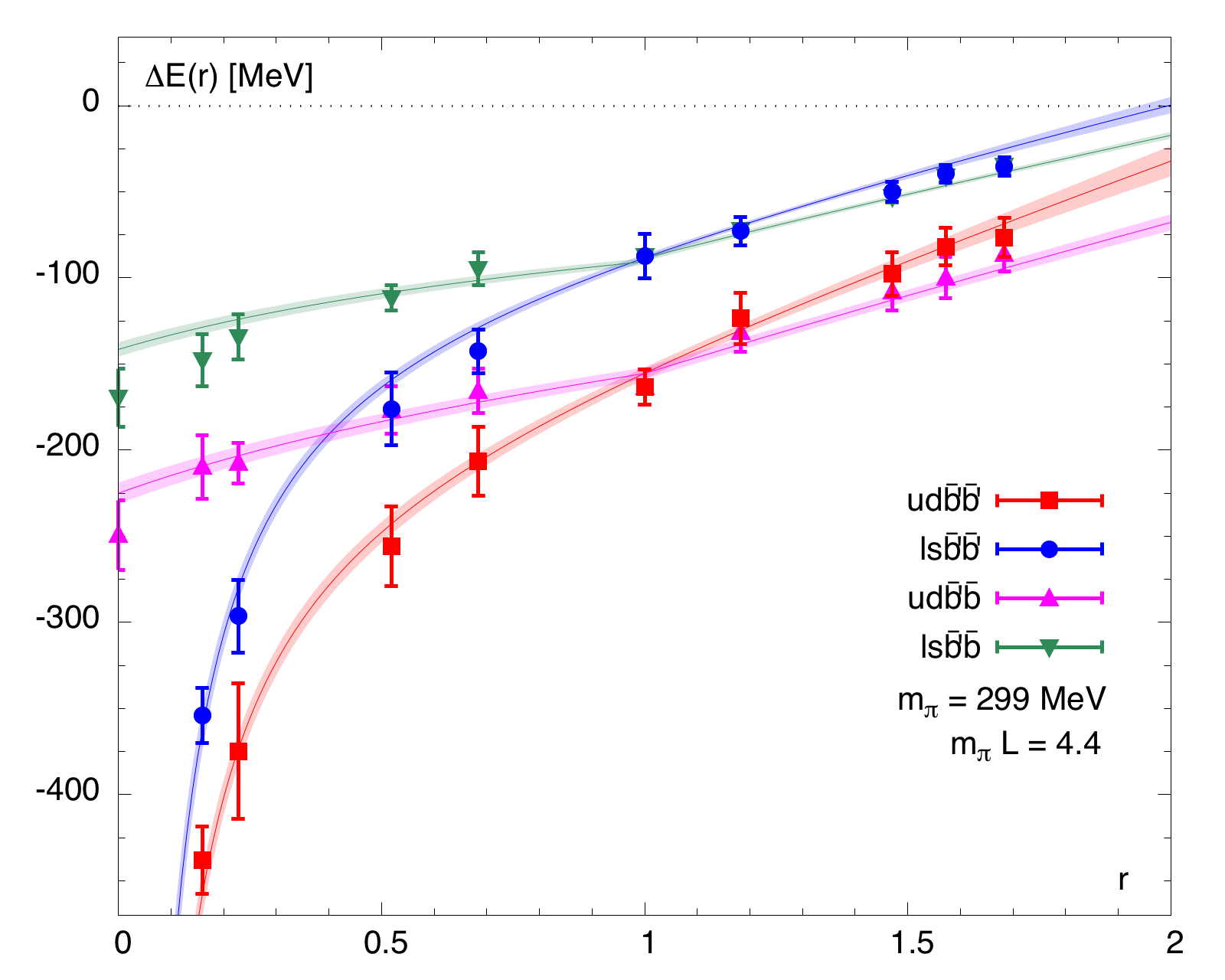}
    \caption{Taken from~\cite{Francis:2018jyb}. The heavy mass dependence of the tetraquark binding energies, plotted against $r=m_b/m_h$. Note that $b'\equiv h$.}
    \label{fig:2018_heavy_mass}
  \end{center}
\end{figure}

The heavy mass dependence of the $J^P=1^+$ binding for $ud\bar{b}\bar{h}$, $ud\bar{h}\bar{h}$, $ls\bar{b}\bar{h}$ and $ls\bar{h}\bar{h}$ is shown in Fig.~\ref{fig:2018_heavy_mass}, with $b'\equiv h$. The coloured bands show phenomenologically motivated fit functions, based on the colour Coulomb and `good-light' diquark interactions discussed above. See~\cite{Francis:2018jyb} for the explicit forms and a detailed discussion.

We see the binding increasing with increasing reduced mass, heading off to $-\infty$ for the $\bar{h}\bar{h}$ tetraquarks, and reaching a finite value, accessible to our simulation via a static propagator, in the $\bar{b}\bar{h}$ cases. This is in keeping with our expectations from phenomenology. We also note that as $r\equiv m_b/m_h\equiv m_b/m_{b'}$ increases for $m_h\to m_c$, the binding becomes much shallower in all cases. The explicit calculation of $ud\bar{b}\bar{c}$\footnote{using a relativistic action for the $c$} reflects this, giving a binding of $38(23)$ MeV.

\subsection{Introduction of box sinks and exploration of other tetraquarks}\label{sec:box}
In the above analyses, propagators were tied together at a single point (a local sink). However, this introduces a bias in the structure of the tetraquark state. The change introduced in~\cite{Hudspith:2020tdf} was the construction of so called `box sinks', the final part of the story leading to our present analysis. This involves summing the propagators over all $N$ spatial points within a chosen radius, $R$,
\begin{equation}
  S^{B,R}(t)= \frac{1}{N}\sum_{r^2\leq R^2}S(x+r,t).
\end{equation}
In the limit $R^2=0$, this is equivalent to a local sink, as before, whilst for large values of $R^2$, we approach a wall sink, where all spatial points on the lattice are summed over.
\begin{figure}
  \begin{center}
    \includegraphics[width=0.48\textwidth]{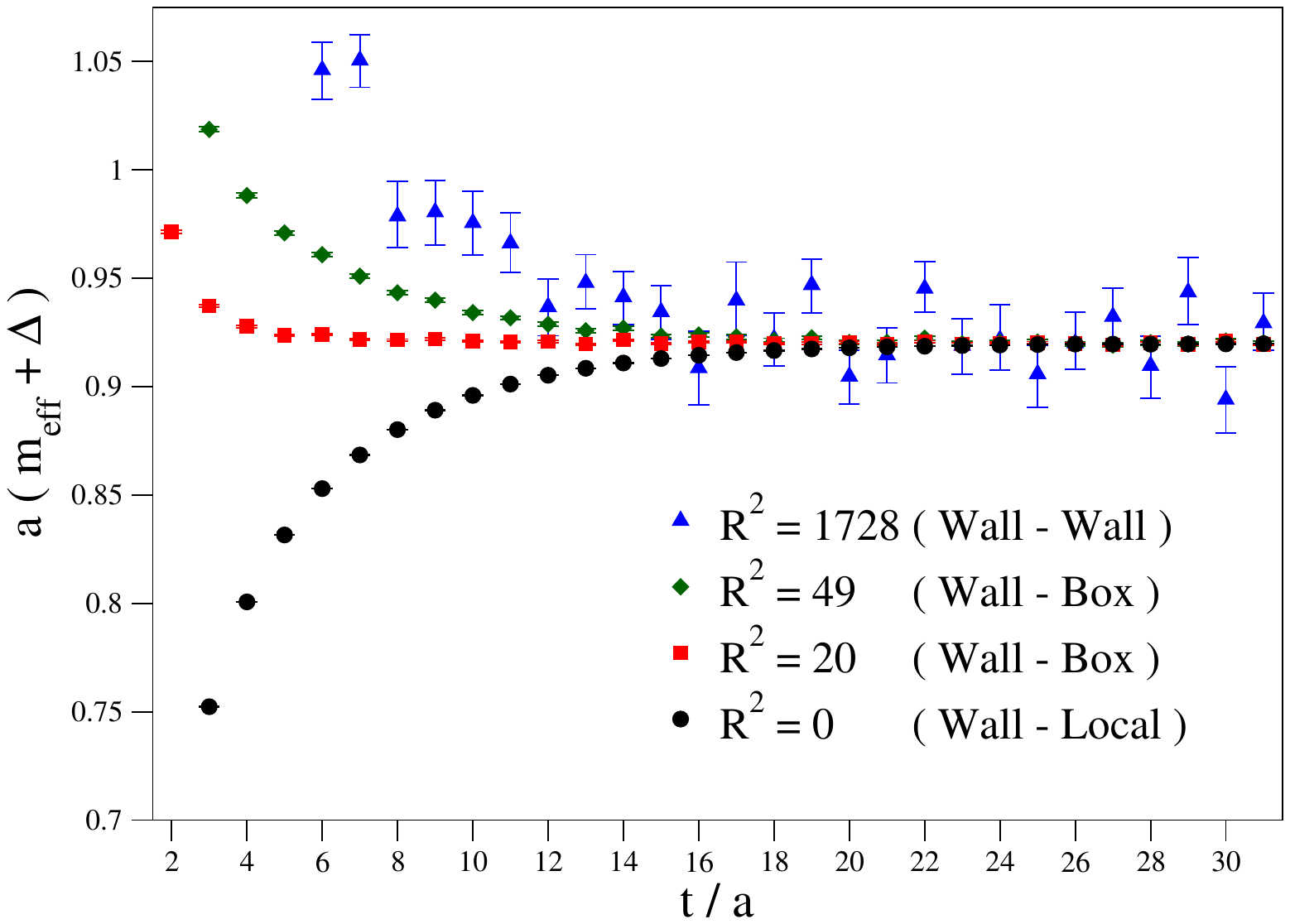}
    \includegraphics[width=0.48\textwidth]{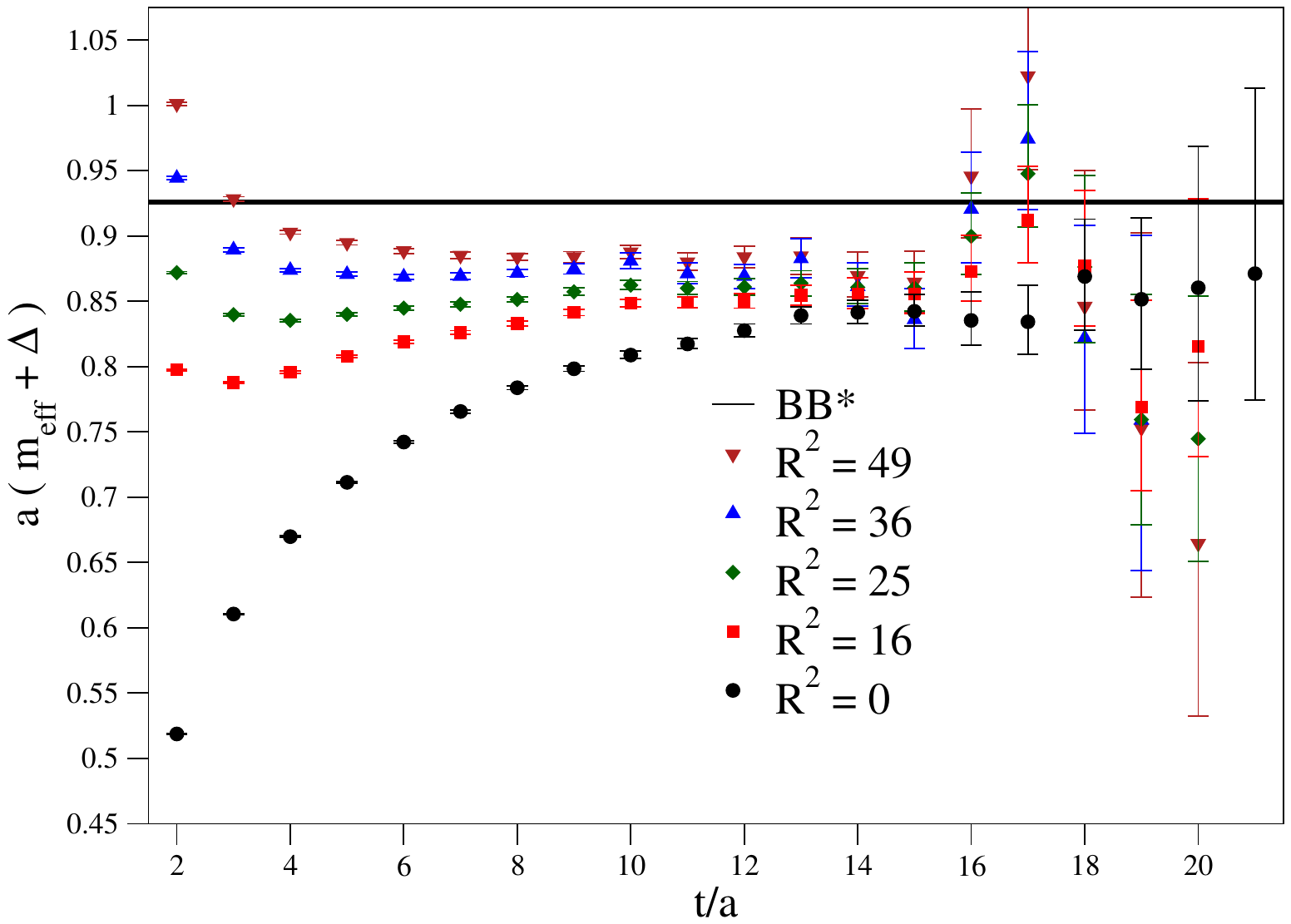}
    \caption{Taken from~\cite{Hudspith:2020tdf}. The effective masses of the $B_c$ meson (left) and $ud\bar{b}\bar{b}$ tetraquark (right), for a variety of $R^2$ choices.}
    \label{fig:2020_box}
  \end{center}
\end{figure}

Fig.~\ref{fig:2020_box} shows the effect of varying $R^2$ on the effective mass, both for a $B_c$ meson, and a $ud\bar{b}\bar{b}$ tetraquark. It's clear that changing the radius of the box sink alters the way that the operator overlaps with excited states, and an optimal choice can improve convergence to the ground state significantly. In addition, including in our analysis multiple radii with different sized excited state contributions gives us a much better handle on systematics associated with radius choices.

Using the box sink method and one pion mass (192 MeV),~\cite{Hudspith:2020tdf} studied $J^P=1^+$ and $J^P=0^+$ tetraquarks for a large number of flavour combinations: $ud\bar{b}\bar{c}$, $ud\bar{b}\bar{s}$, $ud\bar{s}\bar{c}$, $ud\bar{b}\bar{b}$, $ls\bar{b}\bar{b}$, $sc\bar{b}\bar{b}$ and $ls\bar{b}\bar{c}$ in much the same manner as has been discussed above (see~\cite{Hudspith:2020tdf} for full details).

The results of this study were a significant reduction in the bindings of the $J^P=1^+$ $ud\bar{b}\bar{b}$ and $ls\bar{b}\bar{b}$ to around 110 and 40 MeV, respectively. This was attributed to a much better convergence to the plateau than had been achievable in~\cite{Francis:2016hui,Francis:2018jyb} without the box sinks, as well as the use of larger volume ensembles. It was also observed that the $ud\bar{b}\bar{c}$ was not deeply bound, but shallow binding could not be ruled out without a finite volume analysis. No binding was found in other channels. 
\section{Current, updated analysis}
Our latest analysis,~\cite{Colquhoun:2024jzh}, takes the box sink method developed in~\cite{Hudspith:2020tdf}, and applies it to the light quark and heavy quark mass dependence studies of~\cite{Francis:2016hui,Francis:2018jyb}. Taking the findings of~\cite{Hudspith:2020tdf}, we focus our study on $J^P=1^+$ and $J^P=0^+$ tetraquarks of the form $ud\bar{h}\bar{h}$, $ud\bar{b}\bar{h}$, $ls\bar{h}\bar{h}$ and $ls\bar{b}\bar{h}$.   

We use the same set of PACS-CS ensembles as above~\cite{PACS-CS:2011ngu}, but include two additional ensembles, at lighter pion mass, which we have generated with a larger spatial volume of $L/a=48$, and hence a larger minimum $M_{\pi}L=3.6$.
\begin{table}[h]
  \begin{center}
  \begin{tabular}{cccccc}
    \hline
    \hline
    Label &$V$ & $\kappa_l$ & $N_{\mathrm{conf}}\times N_{\mathrm{src}}$ &$aM_{\pi}$& $M_{\pi} L$ \\
    \hline
    E1 & $32^3\times 64$ & $0.13700$ &  $399\times4$ &0.32205(18)& 10.3 \\
    E2 & $32^3\times 64$ & $0.13727$ &  $400\times4$ &0.26193(19)& 8.4 \\
    E3 & $32^3\times 64$ & $0.13754$ &  $400\times4$ &0.18960(29)& 6.1 \\
    E5 & $32^3\times 64$ & $0.13770$ &  $800\times4$ &0.13622(27)& 4.4 \\
    E7 & $48^3\times 64$ & $0.13777$ &  $94\times8$ &0.08719(47)& 4.2 \\
    E9 & $48^3\times 64$ & $0.13779$ &  $88\times4$  &0.07536(58)& 3.6 \\
    \hline
    \hline
  \end{tabular}
  \end{center}
  \caption{The ensembles used. Ensemble labels and volume in lattice units, $V$, are given, as well as $\kappa_{l}$ values, the statistics in terms of configurations and sources per configuration, the pion mass in lattice units ($aM_{\pi}$) and the $aM_{\pi}L$, where $L$ is the spatial extent in lattice units. $a^{-1}=2.194(10)\;\mathrm{GeV}$ from the $\Omega$ mass at the physical point~\cite{PACS-CS:2011ngu}.}
  \label{ensembledetails}
\end{table}
Details of these ensembles are given in Tab.~\ref{ensembledetails}, and further details about their generation can be found in~\cite{Colquhoun:2024jzh}.

As well as the expanded range of six $M_{\pi}$ values for the light mass dependence part of the study, we employ nine $am_h$ values for our heavy mass dependence study, on the E5 ensemble. On each ensemble, we use four\footnote{Two in the case of E7.} different box sink radii, to ensure good resolution of the ground state, as discussed in Sec~\ref{sec:box}. As in~\cite{Hudspith:2020tdf}, we use up to 36 different combinations of source and sink operators.

Instead of using a GEVP to extract the binding energies, we use correlated, multi-exponential fits to both the tetraquark and threshold meson states of the form,
\begin{equation}\label{C2fitform}
  C_2^{\mathcal{O}_{\mathrm{src}}^i\mathcal{O}_{\mathrm{snk}}^j}(t) = \sum_{n=0}^{N}a_n^{\mathrm{src}_i}a_n^{\mathrm{snk}_j}(e^{-E_nt} \pm e^{-E_n(T-t)}),
\end{equation}
where $T=64$ is the (periodic) time dimension of the lattice.

We obtain correlated values for the ground state energies $E_0$. As discussed, our data contains different source and sink combinations, from both the operator choices and box sink radii. These translate into different values for the amplitudes $a_n^{\mathrm{src}/\mathrm{snk}}$, with a common set of energy levels $E_n$.

These Bayesian fits are performed using the \textit{corrfitter}, \textit{lsqfit} and \textit{gvar} python packages~\cite{peter_lepage_2021_5733391,peter_lepage_2023_7931361,peter_lepage_2023_8025535}, and we conduct a rigorous stability analysis to ensure that we are fitting accurately. This is discussed in more detail in~\cite{Colquhoun:2024jzh}. These fits offer a different but equivalent approach to the GEVP analysis, which may be beneficial in the case where the excited states in the spectrum are closely spaced.  

%
\begin{figure}
  \begin{center}
    \includegraphics[width=0.48\textwidth]{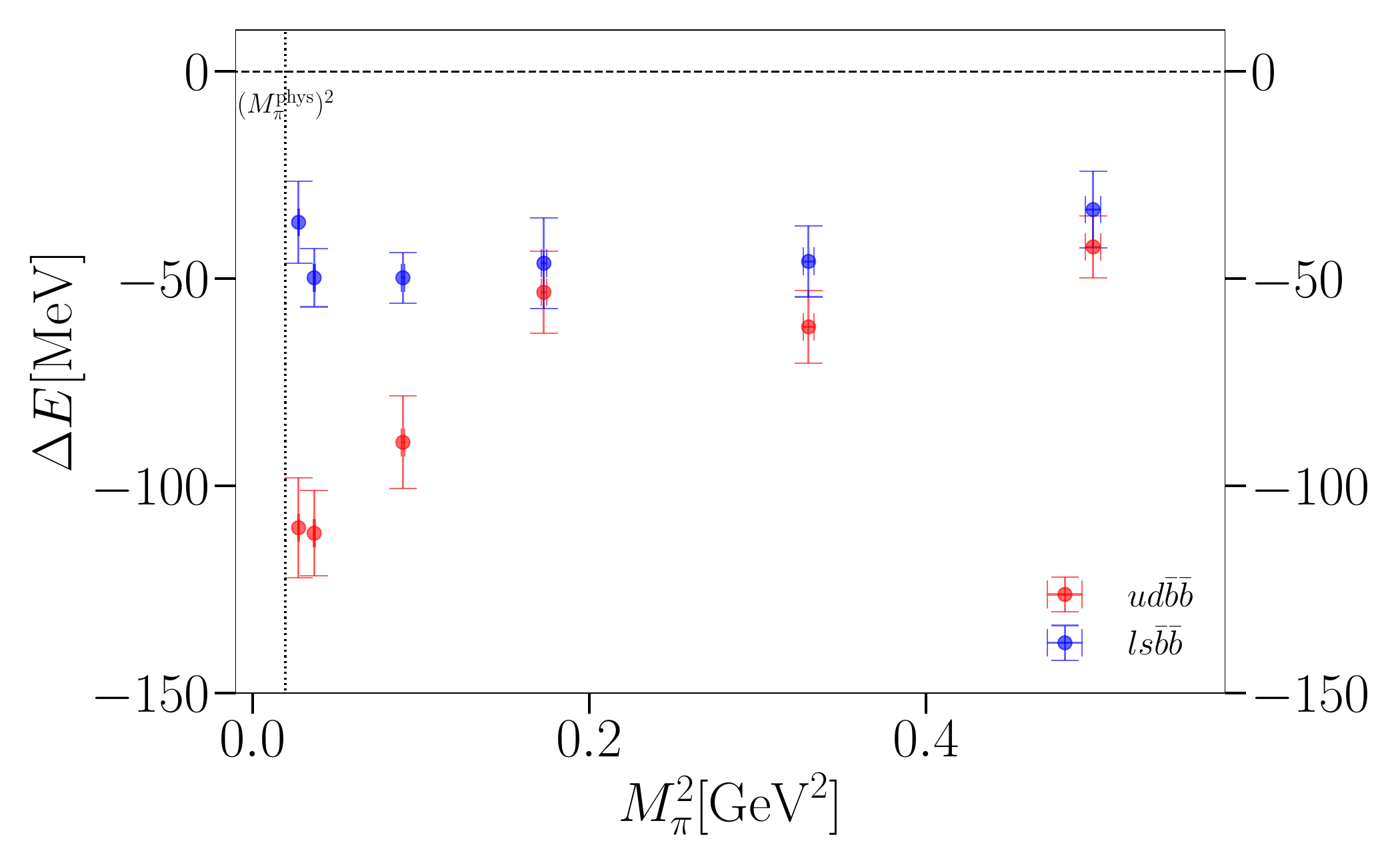}
    \includegraphics[width=0.48\textwidth]{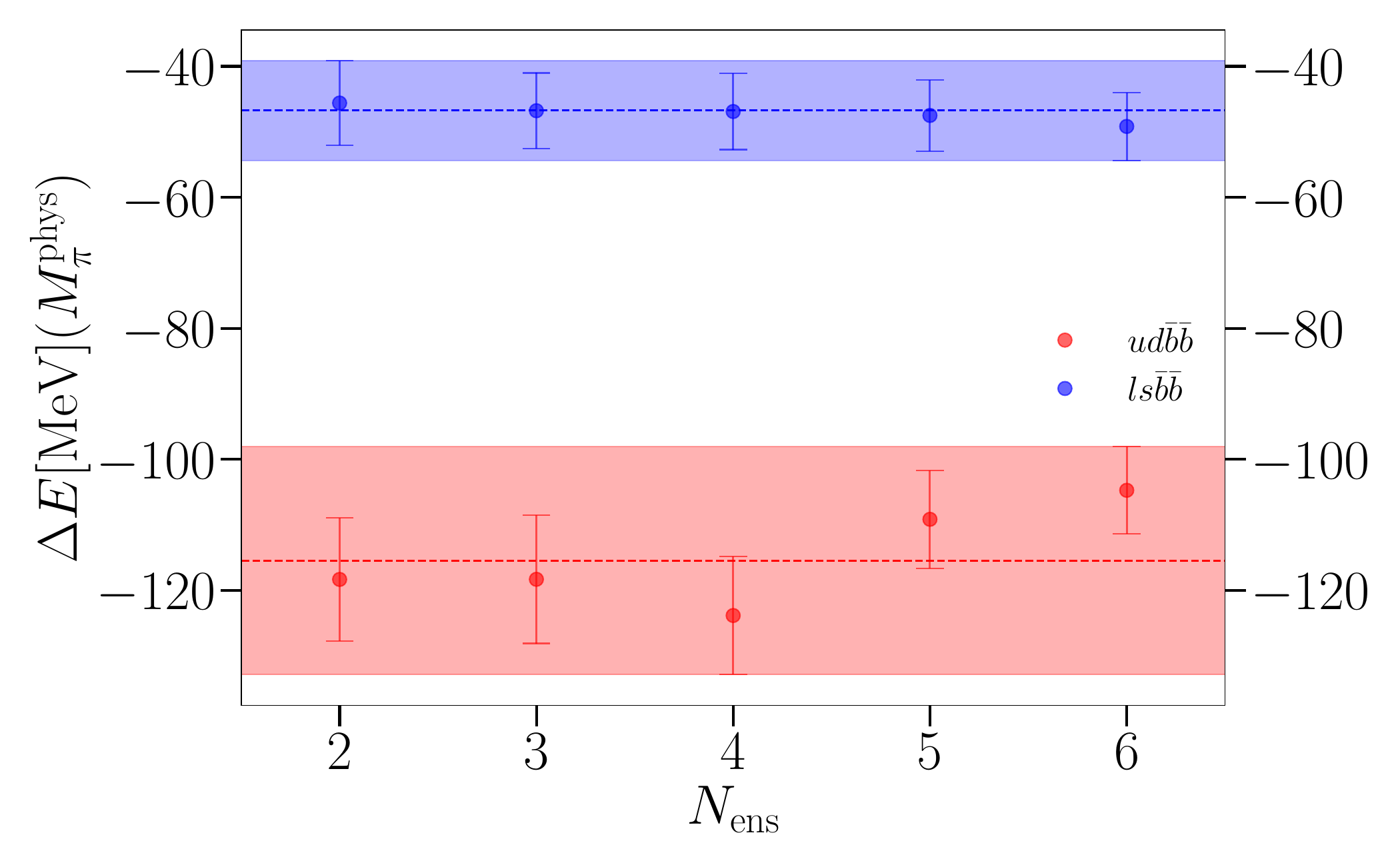}
    \caption{New results from~\cite{Colquhoun:2024jzh}. Left: the updated version of Fig.~\ref{fig:2016_chiral}. Right: the result at physical $M_{\pi}$ when $N_{\mathrm{ens}}$ data points are included in the fit, with the coloured bands giving our final results. }
    \label{fig:2024_chiral}
  \end{center}
\end{figure}
\begin{figure}
  \begin{center}
    \includegraphics[width=0.48\textwidth]{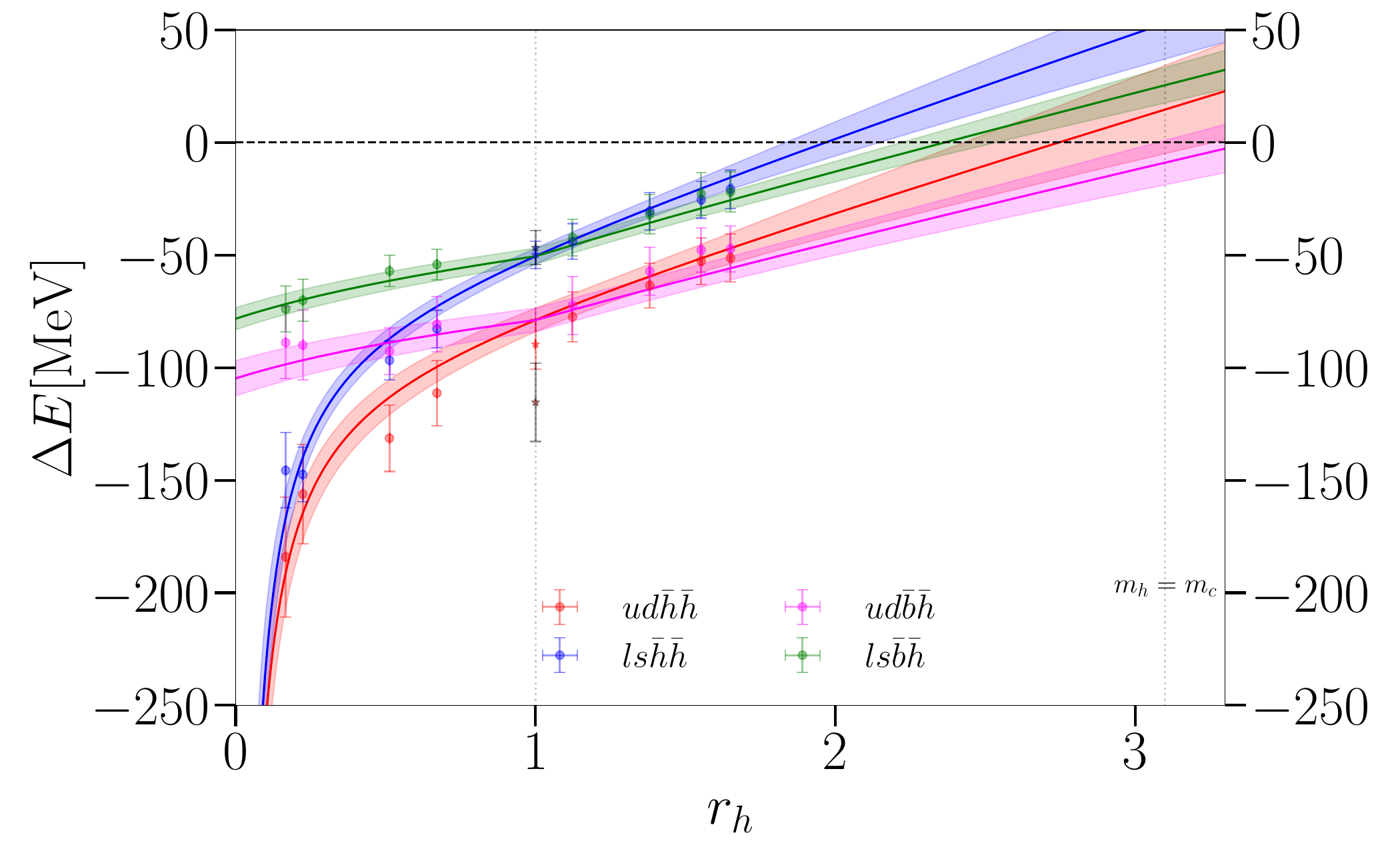}
    \includegraphics[width=0.48\textwidth]{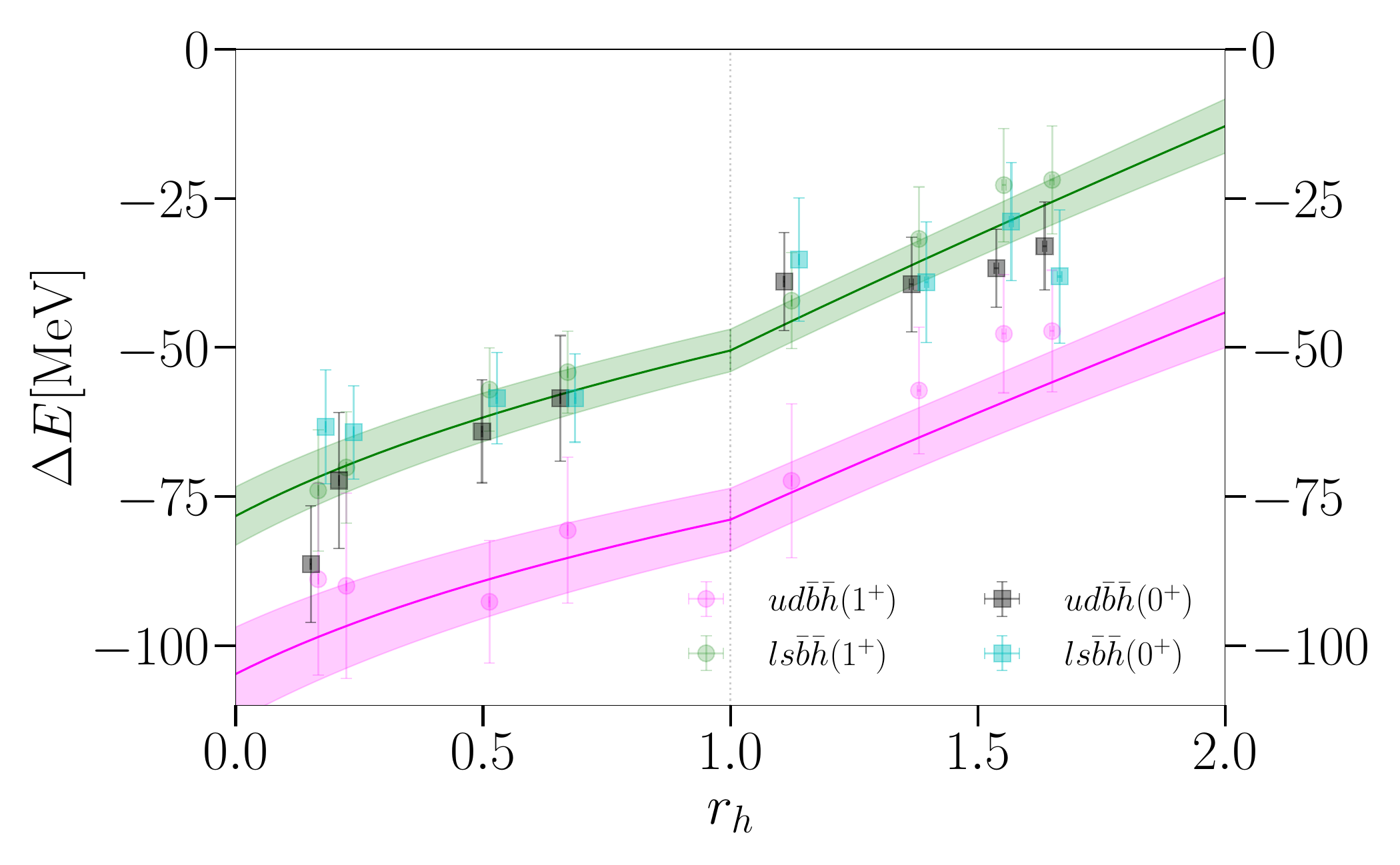}
    \caption{New results from~\cite{Colquhoun:2024jzh}. Left: the updated version of Fig.~\ref{fig:2018_heavy_mass}, showing the heavy mass dependence of $J^P=1^+$ tetraquark binding. Black stars indicating the physical pion mass results from Fig.~\ref{fig:2024_chiral}. Right: a comparison of $J^P=1^+$ and $J^P=0^+$ bindings for $\bar{b}\bar{h}$ type tetraquarks. }
    \label{fig:2024_mass}
  \end{center}
\end{figure}
The results of our light mass dependence study are shown in Fig.~\ref{fig:2024_chiral}, which should be compared with the previous analysis Fig.~\ref{fig:2016_chiral}. The left panel shows the six pion mass values, with the lightest two coming from our new larger $M_{\pi}L$ ensembles. As before, it seems $ud\bar{b}\bar{b}$ is much more sensitive to the light mass than $ls\bar{b}\bar{b}$. We include much larger $M_{\pi}$ values than in the previous study, and it's not necessarily valid to extend the linear fit form $A+BM_{\pi}^2$ to include these large pion masses. For this reason, we perform a fit to increasing numbers of data points, from the leftmost two, up to all six, and plot the resulting $M_{\pi}^{\mathrm{phys}}$ extrapolation in the right hand pane of Fig.~\ref{fig:2024_chiral}. We then take as our final result the band which covers the spread of these values with their respective error bars. Whilst we don't perform a full finite volume analysis, exponential FV effects are included in this fit (see~\cite{Colquhoun:2024jzh}) contributing an increase in binding of about $0.4\sigma$ in $ud\bar{b}\bar{b}$, with no meaningful effect on $ls\bar{b}\bar{b}$. These final results are given by,
\begin{align}\label{eq:phys_bindings1}
  \Delta E_{ud\bar{b}\bar{b}}(M_{\pi}^{\mathrm{phys}}) &= -115(17)~\mathrm{MeV} \\\label{eq:phys_bindings2}
  \Delta E_{ls\bar{b}\bar{b}}(M_{\pi}^{\mathrm{phys}}) &= -46.7(7.6)~\mathrm{MeV},
\end{align}
which represent somewhat shallower binding than in our previous analyses. 

The variation of the binding with heavy masses dependence is show in Fig.~\ref{fig:2024_mass}, which should be compared with Fig.~\ref{fig:2018_heavy_mass}. The left hand pane shows the phenomenologically motivated fit to the $J^P=1^+$ data, again showing a good fit. On the right, we compare $J^P=0^+$ and $J^P=1^+$ data for the $ud\bar{b}\bar{h}$ and $ls\bar{b}\bar{h}$ cases. We see that the $ls\bar{b}\bar{h}$ $0^+$ and $1^+$ give similar results, whilst the $ud\bar{b}\bar{h}$ case shows a relatively consistent difference of order 20 MeV.
\section{Conclusions}
The introduction of box sinks, as well as two new, larger lattice volumes, more pion masses and multi-exponential fits have allowed us to build on the previous work in~\cite{Francis:2016hui,Francis:2018jyb,Hudspith:2020tdf}. We conducted studies of the light and heavy mass dependence of the $J^P=1^+$ $ud\bar{b}\bar{b}$, $ud\bar{b}\bar{h}$, $ls\bar{b}\bar{b}$ and $ls\bar{b}\bar{h}$ tetraquark binding, as well as a comparison with the $J^P=0^+$ $ud\bar{b}\bar{h}$, and $ls\bar{b}\bar{h}$ case.

We find bindings at physical pion masses which are reduced with respect to our previous studies, and fall in line with a general trend towards shallower bindings in recent analyses~\cite{Junnarkar:2018twb,Leskovec:2019ioa,Hudspith:2023loy,Aoki:2023nzp,Alexandrou:2024iwi}.

Whilst our heavy mass dependence analysis does not extend down to the charm mass, we can extrapolate by eye to see that, taking into effect the light mass dependence, it is roughly compatible with the confirmed existence of the $1^+$ $T_{cc}$ just below threshold~\cite{LHCb:2021vvq,LHCb:2021auc}. A similar extrapolation suggests that the $1^+$ $ud\bar{b}\bar{c}$ could also be bound, with shallow enough binding to permit an electromagnetic decay. If the $J^P=0^+$ state is unbound, then the $1^+$ state should have a near 100\% branching fraction to $B\bar{D}\gamma$, producing highly collimated $B$-$\bar{D}$ pairs in the lab frame.
Even if the $J^P=0^+$ state is bound, the $1^+$ state will decay electromagnetically to a soft photon plus the $0^+$ state as well as to $B\bar{D}\gamma$, with a sizeable branching fraction. This $B\bar{D}\gamma$ branch will have the same potentially useful, statistically enhanced signal as it would if the $0^+$ state were unbound.

\bibliographystyle{JHEP}
\bibliography{my_bib}{}


\end{document}